\documentclass[useAMS,usenatbib,referee]{mn2e}
\usepackage{epsfig}
\usepackage{graphicx}
\usepackage{lscape}
\usepackage{color}
\def \vec#1{\mbox{\boldmath$#1$}}
\title[Model on pulsed GeV radiation from magnetars]
{Model on pulsed GeV  radiation from  magnetars}
\author[J. Takata et al.]{J. Takata \thanks{takata@hku.hk}, Y. Wang \thanks{yuwang@hku.hk}, E.M.H. Wu \thanks{wmheric@gmail.com}, 
and  K.S. Cheng\thanks{hrspksc@hkucc.hku.hk}
\\
Department of Physics, University of Hong Kong,
Pokfulam Road, Hong Kong
} 
\begin{document}

\date{}

\pagerange{\pageref{firstpage}--\pageref{lastpage}} \pubyear{2010}

\maketitle

\label{firstpage}

\begin{abstract}
We discuss a possible scenario for radiation mechanism of 
 pulsed GeV $\gamma$-rays from  magnetars. 
 The magnetars  have  shown frequent X-ray bursts, which would
 be triggered  by crust fractures and could release the 
 energy of order of $E_{tot}\sim 10^{41-42} {\rm erg}$. 
If the location of the crust  cracking of the magnetic field  
is close to the magnetic pole,  part of the released energy 
may excite  the Alfv$\acute{\rm e}$n wave 
that can propagate into outer magnetosphere. 
The oscillation of the magnetic field induces the available potential 
drop $\delta\Phi_p\sim 10^{15}$~Volts,
 which can accelerate the electrons and/or 
positrons to the Lorentz 
factor $\Gamma\sim 10^{7}$ in the outer magnetosphere. 
The curvature radiation process at 
outer magnetosphere can produce GeV $\gamma$-rays. If the radiation process 
is occurred above $r\sim 5\times 10^7$~cm from the stellar surface, 
the emitted  GeV $\gamma$-rays can escape from the pair-creation 
process with the 
X-rays and/or the magnetic field.  The expected luminosity of the GeV
 emissions is order of $L_{\gamma}\le 10^{35}~{\rm erg~s^{-1}}$,  
and the radiation process will 
last for a temporal scale of years.  
The expected pulse profiles have a broad  shape  with sometimes sharp  peaks. 
We apply the model to AXP 1E~2259+586.
\end{abstract}


\section{Introduction}
\label{intro}

Soft Gamma-ray Repeaters  and Anomalous X-ray Pulsars (AXPs) have been 
discussed in terms of  ``magnetar'' model, in which the activities of 
the   neutron star are powered by the dissipation of the extremely strong 
magnetic field (Thompson \& Duncan 1995; Thompson, Lyutikov 
\& Kulkarni 2002; Woods \& Thompson 2006; Mereghetti 2008). 
The magnetar's emissions mainly appear in the X-ray bands, 
which are described by the blackbody component 
(with hard tail) below 10~keV plus a hard power law 
component above 10~keV (Kuiper et al. 2006). 
The power law components are often explained by the resonant Compton 
scattering process of the mildly 
relativistic electrons and/or positrons below 10~keV 
(Fern$\acute{\rm a}$ndez \& Thompson 2007; Rea et al. 2008)
 and of the relativistic pairs above 10~keV
 (Beloborodov \& Thompson 2007; Baring \& Harding 2007; Beloborodov 2012), 
respectively.

Although no pulsed GeV emissions from the magnetars have been confirmed, 
it has not been  conclusive  that the magnetars are intrinsically 
 dark in the GeV radiation or present sensitivity  of the $Fermi$ telescope 
is not enough to detect any  pulsed GeV emission from magnetars. 
 It has been argued that the rotation powered activities 
of the magnetars can produce 
the $\gamma$-rays from e.g. the outer gap accelerator (Cheng \& Zhang, 2001). 
However, because temperature of the 
surface  X-ray emissions of the magnetars
is $kT\sim 0.5$keV, which is much higher than the typical 
surface temperature $kT\sim 0.1$~keV of the young pulsars, 
the size of the outer gap, and resultant power of the $\gamma$-ray emissions
  will  be relatively smaller than those of the canonical $\gamma$-ray pulsars,
 indicating less possibility  for the detection of 
 the pulsed $\gamma$-ray emissions from the outer gap of the magnetar. 

Recently, GeV $\gamma$-rays from the supernova remnant (SNR) CTB~109, 
which will 
be associated with AXP~1E~2259+586, have been found in data of 
$Fermi$ $\gamma$-ray telescope. 
 (Castro et al. 2012). Although the emissions from 
SNR CTB~109 has been suggested (Castro et al. 2012), the origin has not 
been confirmed yet. Because the  future $Fermi$ observations will allow us 
 a more deep search of the GeV emissions from the magnetars,
 it will be worth to discuss the possible mechanism  
of GeV $\gamma$-ray emissions in  the magnetosphere of the magnetar. 

The magnetar's X-ray/$\gamma$-ray radiations powered by the magnetic energy 
 have been discussed with the emission process near the stellar 
surface $r<10^{7}$~cm (e.g. Beloborodov 2012). In such a case, the GeV 
$\gamma$-rays cannot escape from the pair-creation processes 
with the X-rays and/or the magnetic field, 
as we will discuss in section~\ref{region}. 
In this paper, therefore, we will discuss a possible scenario for the GeV 
emissions process powered by the  magnetic energy in the outer magnetosphere. 
 In section~\ref{region}, we will discuss the pair-creation processes  
of the GeV $\gamma$-rays and critical radial distance above which the produced 
GeV $\gamma$-rays can escape from the pair-creation processes. In 
section~\ref{model}, we will argue
 that Alfv$\acute{\rm e}$n wave will be excited
 by  crust cracking of the magnetic field
 and carries the released energy into outer 
magnetosphere along the background magnetic field line. We will discuss 
the possible acceleration mechanism due to the propagation of the  Alfv$\acute{\rm e}$n wave in the outer magnetosphere.  We will also estimate the 
Lorentz factor of the accelerated particles and the typical energy of 
the curvature radiation process. We will calculate 
the  expected pulse profiles in section~\ref{profile} and 
 $\gamma$-ray spectra in section~\ref{spectrum}. In  section~\ref{discuss}, 
after brief summary of the results, we will discuss  the possibility of 
the detection for the pulsed GeV radiation from the magnetars.

\section{Pair-creation processes and GeV $\gamma$-ray emission region}
\label{region}
The ultra-strong surface magnetic field of the magnetars permits 
one-photon pair-creation process, 
$\gamma+B\rightarrow e^++e^-$ (Harding \& Lai 2006).  
The condition of the magnetic pair-creation 
process occurred at a radial distance $r$ from the neutron star 
may be written down as (Ruderman \& Sutherland 1975)
\begin{equation}
\frac{E_{\gamma}}{2m_{\rm e}c^2}\frac{B_d(r)\sin\theta_{\rm kB}}{B_{\rm c}}\ge \chi ~~{\rm for}~B_d(r)/B_{\rm c}<0.1,
\label{magcon}
\end{equation}
where $\chi \sim0.1$, $E_{\gamma}$ is photon's energy,
 $B_d(r)$ is the global dipole magnetic 
field at $r$,  and $B_{\rm c}=4.4\times 10^{13}$~G is the 
critical magnetic field strength. In addition,  $m_{\rm e}c^2$ is the electron rest mass energy and  $\theta_{\rm kB}$ is the angle between the magnetic 
field and the propagation direction 
of the photon.   Using the dipole magnetic field, 
we estimate the radial distance, above which  GeV $\gamma$-rays 
can  escape from  the magnetic pair-creation process, 
\begin{equation}
r_{\rm {m}}=3\times 10^7\left(\frac{\chi}{0.1}\right)^{-1/3}
\left(\frac{E_{\gamma}}{\rm{3GeV}}\right)^{1/3}
\left(\frac{B_{\rm d}}{B_{\rm c}}\right)^{1/3}\sin^{1/3}\theta_{\rm kB}~\rm{cm},
\label{rm}.
\end{equation}
 The high-energy photons may be emitted along 
the magnetic field line by the relativistic particles, that is $\theta_{kB}\sim 0$, and  hence they  can initially propagate into the magnetosphere. 
However,  because the field lines have a curvature 
and because the field lines are
 co-rotating with the magnetar, the collision angle increases 
as the photons propagate from the emission point.  
Under the approximation of the concentric
 circles of the magnetic field line,  the collision angle $\theta_{kB}$ 
 will develop as $\sin\theta_{kB}\sim s/\sqrt{R^2_c+s^2}$, 
where $s$ is the propagation distance from the emission point 
and $R_c$ is the curvature 
radius of the magnetic field line. Hence,  the photons emitted 
blow $r\sim r_m$ will be converted into pairs after 
they propagate $s\sim R_c$, where $\sin\theta_{kB}$ cannot be small. 

The GeV $\gamma$-rays  are also subject to the two photon pair-creation process 
with  background soft X-ray field.   Using typical observed luminosity 
$L_{X}\sim 10^{35}~\mathrm{erg~s^{-1}}$ and temperature  $kT\sim 0.5$~keV (e.g. 
Kuiper et al. 2006; den Hartog et al. 2008; Enoto et al. 2010), the number 
density can be estimated as
 $n_X\sim L_{\rm X}/(4\pi r^2c E_{\rm X})\sim 3\times 10^{20}(L_{\rm X}/10^{35}
\mathrm{erg~s^{-1}})(E_{\rm X}/0.5{\rm keV})^{-1}$ $(r/10^6\rm {cm})^{-2}
\mathrm{cm^{-3}}$. The optical depth $\tau\sim n_X\sigma_{\gamma\gamma}r$, where 
$\sigma_{\gamma\gamma} \sim \sigma_{T}/3$ is the pair-creation cross section 
with $\sigma_T$ being the Thomson cross section, is below  unity 
if GeV $\gamma$-rays  are emitted above  the radiation distance
\begin{equation}
r_{\rm {p}}=7\times 10^7\left(\frac{L_{\rm X}}{10^{35}~\mathrm{erg s^{-1}}}\right)
\left(\frac{E_{\rm X}}{0.5\rm keV}\right)~\rm cm.
\label{rp}
\end{equation}
Equations~(\ref{rm}) and~(\ref{rp}) imply that 
the critical radius $r_c$ above which the produced 
 GeV $\gamma$-rays can escape 
from the pair-creation processes is  order of $r_c\sim 5\times 10^{7}$~cm.

\section{Emission Model}
\label{model}
\subsection{The Alfv$\acute{\rm e}$n wave  excited 
by the crust fractures}
\label{alfven}

Woods et al. (2005) proposed that there are two distinct types of magnetar's 
bursts, which were named Type~A and Type~B. Type~A bursts are frequently  
 seen in SGR bursts, in that they are uncorrelated with pulse phase and 
the energy emitted during primary burst peak is larger than the tail energy. 
Type~B bursts, on the other hands, are correlated with the 
pulse phase and the energy of primary burst peak is smaller 
than the tail energy (see also Kaspi 2007; Scholz \& Kaspi 2011; 
Dib et al. 2012). Woods et al. (2005) also speculated that the Type~A and Type~B bursts are triggered by the magnetospheric reconnection (Lyutikov 2003)
 and by the crust fractures (Thompson \& Duncan 1995), respectively. 
 Furthermore,  several magnetars sometimes show   timing  glitches that 
accompany  the radiative outbursts  
(Woods et al. 2004;\.{I}\c{c}dem et al. 2012; 
Dib et al. 2012; Pons \& Rea 2012). In 2002 giant glitch 
of  AXP 1E~2259+586, for example, the X-ray outburst were consisted of 
a rapidly decay emission in the first few hours with a released 
energy $\sim 10^{38}\rm{ergs}$  and a slow decay emission lasting 
several years with a released energy $>10^{41}~\rm ergs$ 
(Woods et al. 2004; Zhu et al. 2008), which are Type~B burst. 
This giant glitch accompanying the radiative outburst is likely 
triggered by the crust fractures, which would simultaneously affect 
both super-fluid core and the magnetosphere, and  will inject  a energy  
$\sim 10^{41-42}$erg  into the magnetosphere (Woods et al. 2004; 
Pons \& Rea 2012).

In this paper, we consider the energy release caused by the crust cracking 
of the strong magnetic field, i.e. the Type~B bursts. 
 We assume that the magnetic field in the crust is deformed away from the equilibrium state by an amount of $\delta B$. The crust fracture will be occurred if the Maxwell stress $\delta B B_{\rm curst}/4\pi$, where  $B_{\rm crust}$ is
 the  equilibrium magnetic field in the crust, exceeds crustal elastic stress.  
The available magnetic energy density when the crust fracture is occurred  may be expressed 
by (Thompson \& Duncan 1995)
\[
\frac{(\delta B)^2}{8\pi}\sim 2\pi \mu^2 \epsilon_y^2B_{\rm crust}^{-2}, 
\]
where $\mu$ is the shear modulus and  $\epsilon_y$ is the yield strain at which 
the crust cracks. The released magnetic energy is therefore estimated as 
\begin{equation}
E_{tot}=2\times 10^{42} \left(\frac{\mu}{10^{30}\rm {erg~cm^{-3}}}\right)^2
\left(\frac{\epsilon_{y}}{2\cdot 10^{-3}}\right)^2\left(\frac{B_{\rm crust}}{10^{14}\rm G}\right)^{-2}\left(\frac{\ell}{10^{5}\rm cm}\right)^3~\rm ergs, 
\label{totene}
\end{equation}
where  $\ell$ is the typical size of the cracked platelet.  Note that with a typical luminosity of the observed X-ray emissions $L_{X}\sim 10^{34-35}\mathrm{erg~s^{-1}}$, the emission will last a temporal scale of years, $\tau=E_{tot}/L_{X}\sim 10^{7-8}$s .

The process of  the $\gamma$-ray emission induced by the neutron star quake 
was  discussed by Blaes et al. (1989), 
who investigated the possibility for the origin 
of $\gamma$-ray bursts. In their model, 
the neutron star quake will excite the oscillation of the magnetic 
field frozen into the star's crust, and  the Alfv$\acute{\rm e}$n waves 
 carrying  the released energy into the magnetosphere. 
 The dissipation of wave energy will  accelerate the electrons and/or positrons,
 which in turn radiate $\gamma$-rays.

In the magnetar model, it can be thought that the Alfv$\acute{\rm e}$n wave 
is excited on  the magnetic field lines anchored on 
the cracked platelet. If location of the cracking is close to the magnetic 
pole, a part of the released energy  $E_{\rm tot}$ can be carried by 
the Alfv$\acute{\rm e}$n waves that propagate 
 into the outer magnetosphere $r\ge r_c\sim  5\times 10^7$~cm, where 
the $\gamma$-rays can escape from the pair-creation processes. 
 Because the Alfv$\acute{\rm e}$n wave propagates along the closed 
magnetic field lines, the wave may bounce many times between the 
footprints of the field lines. The time-averaged amplitude of the magnetic 
field corresponding to the Alfv$\acute{\rm e}$n 
wave energy density, $\sim E_{tot}/R_c^3$  may be described  as 
\begin{equation}
\delta B(R_s)\sim 5\times 10^{10}
\left(\frac{E_{tot}}{10^{42}\rm erg}\right)^{1/2}
\left(\frac{R_c}{2\cdot 10^{7}\rm cm}\right)^{-3/2}~\rm G
\label{pertmag}
\end{equation}
near the stellar surface. 

As the Alfv$\acute{\rm e}$n wave  propagates from the 
stellar surface into outer magnetosphere,  the amplitude  will  evolve 
as $\delta B(r)\propto A^{-1/2}(r)$, 
where $A(r)$ is the cross section  of the oscillating magnetic flux tube. 
Because the background dipole field is proportional to 
 $B_d(r)\propto A^{-1}(r)$,  the ratio of the perturbed and background magnetic fields evolves with the radial distance as   
\begin{equation}
\frac{\delta B(r)}{B_d(r)}=\frac{\delta B(R_s)}{B_d(R_s)}
\left(\frac{B_d(R_s)}{B_d(r)}\right)^{1/2}\sim 10^{-3}
\left(\frac{B_d(R_s)}{B_d(r)}\right)^{1/2}.
\label{ratio}
\end{equation}

 In the magnetosphere of magnetar, the 
 speed of the Alfv$\acute{\rm e}$n wave exceeds the speed of light, that is, 
\begin{equation}
v_A\equiv \frac{B}{\sqrt{4\pi\rho}}>>,c
\end{equation}
where the $\rho$ is the mass density. In the limit of $v_A\gg c$, 
 we can see that the required current density ($i_a$) to support the propagation of 
 the Alfv$\acute{\rm e}$ wave at distance $r$ is  in order of 
(see Appendix)
\begin{equation}
\frac{i_a(r)}{\kappa i_{GJ}(r)}\sim \frac{\omega\beta_{co}(r)\delta B(r)}{\kappa \Omega B_d(r)}
\sim 10\beta_{co} \left(\frac{\omega}{10^{4}\rm Hz}\right)
\left(\frac{\Omega}{1\rm Hz}\right)^{-1}
\left(\frac{\delta B(R_{\rm s})/B_d(R_{\rm s})}{10^{-3}}\right)
\left(\frac{B_d(R_{\rm s})}{B_d(r)}\right)^{1/2},
\label{current}
\end{equation}
where $i_{GJ}=\Omega B/2\pi$ is the Goldreich-Julian current density, 
and  $\omega$ is the typical shear frequency of the 
wave, which may be estimated as 
 $\omega\sim \rho_{\rm crust}^{-1/2}\mu^{1/2}\ell^{-1}\sim 10^{4}~{\rm Hz}$ with 
$\rho_{\rm crust}\sim 5\times 10^{11} {\rm g~cm^{-3}}$ being the typical 
 mass density in the crust. 

We can see that in the limit of $v_A\gg c$, the induced electric field 
is same order of
 magnitude as the perturbed magnetic field, $\delta E\sim \delta B$
 (see Appendix). Hence, the maximum magnitude of the induced
 electric potential due to perturbation  of the magnetic field lines 
may be  estimated as 
\begin{equation}
\delta \Phi_p\sim \delta\ell \times \delta B =1.5\times 10^{15}
\left(\frac{\delta B(R_s)/B_{\rm d}(R_s)}{10^{-3}}\right)
\left(\frac{\delta B(R_s)}{5\cdot 10^{10}\rm G}\right)
\left(\frac{\ell }{10^{5}\rm cm}\right)
~\rm Volts, 
\label{poten}
\end{equation}
where $\delta\ell$ is the displacement of footprints of the oscillating magnetic
 lines  and was  estimated as 
\begin{equation}
\delta\ell \sim \frac{\delta B(R_s)}{B_{\rm{d}}}\ell
\sim 10^{2}
\left(\frac{\delta B(R_s)/B_{\rm d}}{10^{-3}}\right)
\left(\frac{\ell}{10^{5}~\rm cm}\right)~{\rm cm}.
\label{dell}
\end{equation} 
The   maximum  radiation power  may be  estimated as 
\begin{equation}
L_{\rm r}\sim \delta \Phi_p\times I
\sim  4\times 10^{35}\left(\frac{\Omega}{1\rm Hz}\right)
\left(\frac{\delta B(R_{\rm s})}{5\cdot 10^{10}\rm G}\right)^2
\left(\frac{\ell}{10^{5}\rm cm}\right)^3\left(\frac{I}{I_{GJ}}\right)~\rm {erg~s^{-1}},
\label{power}
\end{equation}
where $I$ is the total current  and $I_{GJ}\sim i_{GJ}\ell^2$

\subsection{Acceleration and GeV emission process}
\label{curv}

\begin{figure}

\includegraphics[height=10cm,width=10cm]{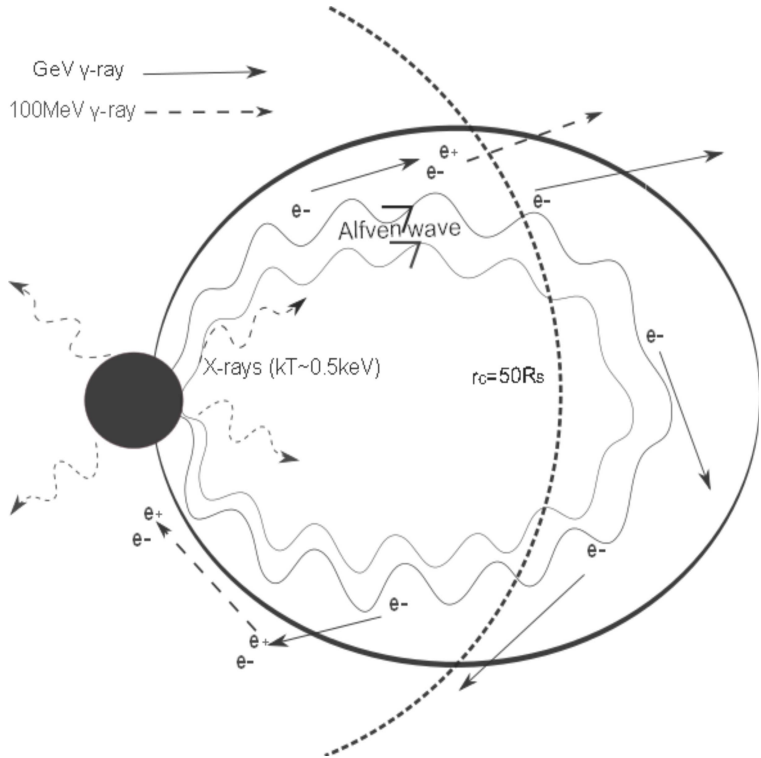}

\caption{Schematic view for the pulsed GeV $\gamma$-ray emissions 
from magnetars.  The figure represents snap shot at which the  
Alfv$\acute{\rm e}$n wave propagates from the north pole to south pole. 
The  GeV $\gamma$-rays produced above 
$r_c\sim 5\times 10^{7}$~cm can escape from the pair-creation process. 
 The created pairs below $r_c$ eventually heat up the stellar surface 
and will contribute to the soft X-ray emissions from the stellar surface. Some 
of outwardly propagating  100~MeV photons emitted via
 the synchrotron radiation of the first generation of  pairs will also 
escape from the pair-creation process, if they  are produced 
beyond  $r_{c1}\sim 10^{7}$cm.}
\label{mag}
\end{figure}

In Figure~\ref{mag}, we  illustrate the schematic view of our model 
for the particle acceleration and GeV emissions in the magnetosphere 
of the magnetar.  
If the magnetic field cracks the platelet near the magnetic axis, 
  the oscillating magnetic flux tube will extend beyond 
$r_c\sim 5\times 10^7$~cm.   We will argue that the particles 
are accelerated by the Alfv$\acute{\rm e}$n wave,
 which is excited by the crust cracking of the magnetic field. 
It will be possible that the energy dissipation 
of the Alfv$\acute{\rm e}$n wave accelerates the electrons and/or positrons 
 to a Lorentz factor above $\Gamma\sim 10^7$. The curvature radiation 
process of the accelerated particle  can produce GeV $\gamma$-rays.
 The GeV $\gamma$-rays emitted above $r_c\sim 5\times 10^7$~cm will escape 
from the pair-creation process and may be   observed as the pulsed emissions.  
 In this section, we will discuss possible 
particle acceleration processes   related to 
 the propagation of the  Alfv$\acute{\rm e}$n wave. 

 First,  if the co-rotation motion is ignored $\vec{v}_{co}=0$, 
the induced electric field $\delta\vec{E}$ and 
the background magnetic field are  a mutually orthogonal 
and hence no electric field along the background magnetic field is induced. 
However, under the presence of the co-rotation motion of the plasmas, such 
orthogonality breaks down, and the electric field and the current along the 
background magnetic field line are induced (Kojima \& Okita, 2004). 
The typical strength  of the 
electric field along the background magnetic field line, $\delta E_{||}$,
 is described as (see Appendix)
\begin{equation}
\delta E_{||}(r)\sim \beta_{co}\delta B(r).
\end{equation}
If the Lorentz factor of particle is determined by the balancing the 
between the electric force and the curvature radiation reaction force, 
we obtain
\begin{equation}
\Gamma(r)\sim \left(\frac{3R_c^2}{2e}\delta E_{||}\right)^{1/4}
\sim 3.5\times 10^7\left(\frac{\delta B(R_s)}{5\cdot10^{10}{\rm G}}\right)^{1/4}
\left(\frac{\beta_{co}}{10^{-3}}\right)^{1/4}
\left(\frac{r}{10^{8}{\rm cm}}\right)^{-3/8}\left(\frac{R_c}{10^8{\rm cm}}
\right)^{1/2},
\end{equation}
where $R_c$ is the curvature radius of the dipole field. We can
 see that the typical energy of the curvature photons 
becomes  several GeV, that is, 
\begin{equation}
E_c=\frac{3}{4\pi}\frac{hc\Gamma^3}{R_c}
\sim 8\left(\frac{\Gamma}{3\cdot 10^7}\right)^3
\left(\frac{R_c}{10^8 \rm cm}\right)^{-1} \rm GeV.
\label{cene}
\end{equation}

Second, the deficiency of the current (or charge) to support the frozen 
in condition, $\vec{E}+\vec{\beta}\times \vec{B}=0$,  could induce the electric 
field along the magnetic field (see also 
Blaes et al. 1989; Fatuzzo \& Melia 1993; Thompson 2006). 
We can see in equation~(\ref{current}) that the required current to support 
the propagation of the Alfv$\acute{\rm e}$n wave easily exceeds  the 
Goldreich-Julian current.  Furthermore, the typical charge density of the 
 unperturbed particles in the closed field region 
will be coincide with  the Goldreich-Julian charge density of the dipole field, 
that is $\rho_{e}=-\vec{\Omega}\cdot\vec{B}_d/2\pi c$.
 If the Alfv$\acute{\rm e}$n wave propagates into outer magnetosphere, however, 
the perturbation of the  magnetic field line changes the 
 local Goldreich-Julian charge density  as 
\begin{equation}
\rho_{GJ,1}=-\frac{\vec{\Omega}\cdot(\vec{B}_{d}+\delta\vec{B})}{2\pi c}
=\rho_{GJ,0}+\delta\rho_{GJ}
\end{equation}
where $\delta\rho_{GJ}\equiv -\vec{\Omega}\cdot\delta\vec{B}_/2\pi$. 
The  electric field along the magnetic field line then will be induced, 
if the primary particles could not supply the required 
charge density $\rho_{GJ,1}$. The particles could
 be accelerated at 
the ``charge-starved'' region up to the Lorentz factor $\Gamma\sim 10^7$, 
such as the particle acceleration process of the $\gamma$-ray 
pulsars (e.g. Wang, Takata \& Cheng, 2010).
 The accelerated particles will  emit 
the $\gamma$-rays via the curvature radiation process. Most of 
the  $\gamma$-rays  produced below $r\sim r_c$ will be converted into  electron
 and positron pairs through the pair-creation processes.  
Because the GeV  $\gamma$-rays  are emitted along the magnetic field lines, 
the pairs are produced at the convex side of the magnetic field line on which 
primary  $\gamma$-rays were emitted (c.f. Figure~\ref{mag}). 
 The multiplicity is roughly  estimated as $\kappa\sim \lambda \times 
e\delta\Phi/\Gamma m_ec^2=2\times 10^3(\Phi/10^{15}{\rm Volt})
(\Gamma/3\cdot 10^7)^{-1}$ with $\lambda\sim r_c/(5\times 10^8)\rm cm \sim 0.1$
 being the fractional power spent  below $r\sim r_c$. A fraction of the 
created  pairs is  probably charge-separated, 
and will produce the required current (\ref{current}) to 
support the propagation of the  Alfv$\acute{\rm e}$n wave.

Finally if the Alfv$\acute{\rm e}$n wave does not efficiently 
dissipate  along the background magnetic field lines, the ratio of the 
amplitude  of the Alfv$\acute{\rm e}$n wave to the dipole field 
will  evolve as $\delta B(r)/B_d(r)\propto 10^{-3} (B_d(R_s)/B_d(r))^{1/2}$,
 as equation~(\ref{ratio}) shows. 
Hence, we find that the fractional perturbation of the magnetic field 
becomes order of unity at the radial  distance  $r\sim 10^8 
[\delta B(R_s))/10^{-3}B_d(R_s)]^{-2/3}$~cm.  Because 
the induced electric field is same order of
 magnitude as the perturbed magnetic field, $\delta E\sim \delta B$, 
the total electric field is of order of the magnetic field $|E|\sim |B|$, 
where conversion from the electromagnetic energy into the particles energy 
could be possible (Beskin \& Rafikov 2000). 
 When the nonlinear term becomes to be
 important, by whatever process, a substantial part of the wave energy is 
probably converted into the electron/positron energy, which in turn 
radiate  the GeV $\gamma$-rays. 

In any acceleration processes process discussed above,  
the typical  Lorentz factor  will be characterized by 
\begin{equation}
\Gamma_{max}\sim \left(\frac{3R_c^2}{2e}E_{||}\right)^{1/4}
=3\times 10^7\left(\frac{R_c}{10^8\rm cm}\right)^{1/4}
\left(\frac{L}{R_c}\right)^{-1/4}
\left(\frac{\delta B(R_{\rm s})}{5\cdot 10^{10}\rm G}\right)^{1/2} \nonumber
\left(\frac{B_{\rm d}(R_s)}{10^{14}\rm G}\right)^{-1/4}
\left(\frac{\ell}{10^5\rm cm}\right)^{1/4},
\label{gmax}
\end{equation}
where we used typical electric field $E_{||}=
\delta \Phi_p/L$ with $L$ being the arc length of the acceleration region 
along the magnetic field line.

\subsection{Incident X-ray/soft $\gamma$-ray  emission}
\label{xemi}
In the present scenario, the relativistic electrons/positrons, 
which were accelerated  in outer magnetosphere, will be return to the stellar surface, because they migrate along the closed magnetic field lines.  
 Below $r_c\sim 5\times 10^7$~cm, most of $\gamma$-rays emitted by incoming particles 
 will be converted into electron/positron pairs 
via the pair-creation process with the soft X-rays and/or magnetic field.
 Those pairs are mainly produced at the convex side of the magnetic 
field lines, on which the primary $\gamma$-rays were emitted. New born pairs 
will loose their energy  via the synchrotron radiation
 and/or the resonant Compton scattering,
 in which the Compton scattering is occurred in the resonant energy
 $\Gamma_{e\pm} E_{X}\sim \hbar eB_d(r)/m_ec$ with $\Gamma_{e\pm}$
 being Lorentz factor of the pairs. 
 Those  photons could   also make new  pairs. This
 cascade process will produce lots of low-energy 
electron/positron pairs near the stellar surface. The incoming pairs 
will carry the energy of $L_{e\pm}\sim \lambda L_{r}\sim 
 4\times 10^{34}\rm {erg~s^{-1}}$.

 The incoming pairs  eventually reach the stellar surface and  
heat up the stellar surface.  The temperature of the heated 
platelet may be $ k_B(L_{e\pm}/\ell^2\sigma_{SB})^{1/4}$$
\sim 2(L_{e\pm}/10^{35}\rm {erg~s^{-1}})^{1/4}$ $(\ell/10^{5}\rm cm)^{-1/2}$ keV,
 with $k_B$ being the Boltzmann constant and  $\sigma_{SB}$ 
Stephan-Boltzmann constant. These hard X-rays emitted from the heated platelet 
may be too high to escape from the resonant Compton scattering near the 
stellar surface.  The scattered X-rays will be redistributed on the stellar surface, and 
emerge as the soft X-ray emission with 
a temperature of $kT_s\sim k(L_{e\pm}/\triangle\Omega R_s^2\sigma_{SB})^{1/4}
\sim  0.6(L_{e\pm}/10^{35}\rm{erg~s^{-1}})\triangle\Omega^{-1/4}$~keV, were 
$\triangle\Omega $ is the solid angle of the emission. 
It is also possible that  because  some fraction of the incoming  
pairs will  be created outside the oscillating magnetic field lines 
that  connect to the cracked platelet, 
the typical size of the heated surface by the  incoming particles
 is larger than $\ell$. In the present scenario, hence, the temperature of 
the intrinsic emissions from the heated surface by the incoming particles will 
be $kT_s\sim 0.5$keV.

For  ``outwardly'' propagating GeV $\gamma$-rays produced 
near but below  $r\sim 5\times 10^7$~cm, the cascade process 
will not develop to reduce the photon energy to $\sim$MeV, and 
stop after producing the first or second generation of pairs. For example, 
the first generation of pairs produced by 
the magnetic pair-creation process will emit the synchrotron photons 
with an energy $\sim E_{\gamma}/20\sim 100$MeV, 
where $E_{\gamma}$ is the primary GeV $\gamma$-ray produced by the 
curvature radiation (Cheng et al. 1998). The outwardly propagating photons with
 $\sim 100$MeV will escape from the 
magnetic pair-creation processes if they are produced beyond 
$r_{c1}\sim 10^7$~cm (c.f. equation~(\ref{magcon})).
 Between $r_{c1}\sim 10^7$cm and $r_c\sim 5\times 10^7$cm, hence, outwardly 
propagating $\sim 100$~MeV photons emitted by the first geration pairs will 
escape from the magnetosphere of the magnetars, and will be contribute to 
the observed spectrum. We expect that the luminosity of this component
 will be $L_{100MeV}\sim \lambda L_{r}$, 
where $\lambda\sim (r_c-r_{c1})/(5\times 10^8{\rm cm})\sim0.1 $.

\subsection{Pulse profiles}
\label{profile}
In this section, we discuss the expected pulse profiles of the 
$\gamma$-ray emissions. We refer 
the previous studies (e.g. Takata, Chang \& Cheng 2007) for 
 the calculation  of the pulse profile. We apply the vacuum rotating 
dipole field as the magnetic field in the magnetosphere. 
We describe the polar angle of the 
magnetic field lines at 
the stellar surface with $a\equiv \theta(\phi)/\theta_p(\phi)$, 
where $\theta_p$ is the  polar angle of the last-open field line 
 and $\phi$ is the magnetic azimuth. The open and closed magnetic field lines 
correspond to  $a<1$ and $a>1$, respectively. Because the polar cap radius is 
only $\sim$1\% of the stellar radius, it is  likely  that the cracking platelet by 
the magnetic field is  occurred at 
the closed field line region. As we discuss in section~\ref{region}, the GeV $\gamma$-rays 
can escape from the pair-creation process if they are produce beyond  
$r\sim 5\times 10^{7}$cm from the stellar surface. In this section,  hence, 
we apply $a=16$, in which the magnetic field lines extend up to $r\sim 5\times 10^8$~cm, 
 as the main emission region.  
The width of the emitting magnetic flux tube in the azimuthal
 direction  is assumed to be  $\delta\phi\sim 1$~radian, because 
the typical size of cracked platelet is  $R_s \theta_c\delta\phi\sim \ell \sim 10^{5}$cm, where $\theta_c\sim \sqrt{R_s/r_c}\sim 0.1$~radian 
is the polar angle of the magnetic field lines that  extend beyond $r\sim r_c\sim 5\times 10^7$cm.

 We assume a constant emissivity along the magnetic field line, and 
at each point we express the  opening angle of the  $\gamma$-ray cone 
as     
\begin{equation}
\theta_{\rm GeV}\sim \frac{\delta B(r)}{B_d(r)}=10^{-3}\left(\frac{r}{R_s}\right)^{3/2},
\end{equation}
which represents the effect of  the oscillation of the 
 magnetic field lines, on which 
the Alfv$\acute{\rm e}$n wave propagates (c.f. equation~(\ref{ratio})). 
The direction of the centre of the radiation cone coincides with 
the direction of the background vacuum dipole field.  
Because the Alfv$\acute{\rm e}$n wave propagates along the closed 
magnetic field lines, the wave may bounce many times between the 
footprints of the field lines.  The crossing time scale of the 
Alfv$\acute{\rm e}$n wave, 
$\tau_c\sim 10^8{\rm cm}/V_A\sim 0.01$s with $V_A\sim c$ being the speed of 
the Alfv$\acute{\rm e}$n wave, is much shorter than the rotation period.  
To calculate the pulse profile, hence,  we take into account the $\gamma$-ray 
emissions from  both  particles  migrating  from north to south poles and 
 from south to north poles. The inwardly propagating
 $\gamma$-rays may be absorbed by the pair-creation process with the background
 X-rays and/or the magnetic field if they pass through near the stellar surface.
To take into account this effect,  we ignore the contribution from 
the inwardly propagating $\gamma$-rays that pass through 
the region $r< 5\times 10^{7}$~cm. 

\begin{figure}
\includegraphics[height=10cm, width=15cm]{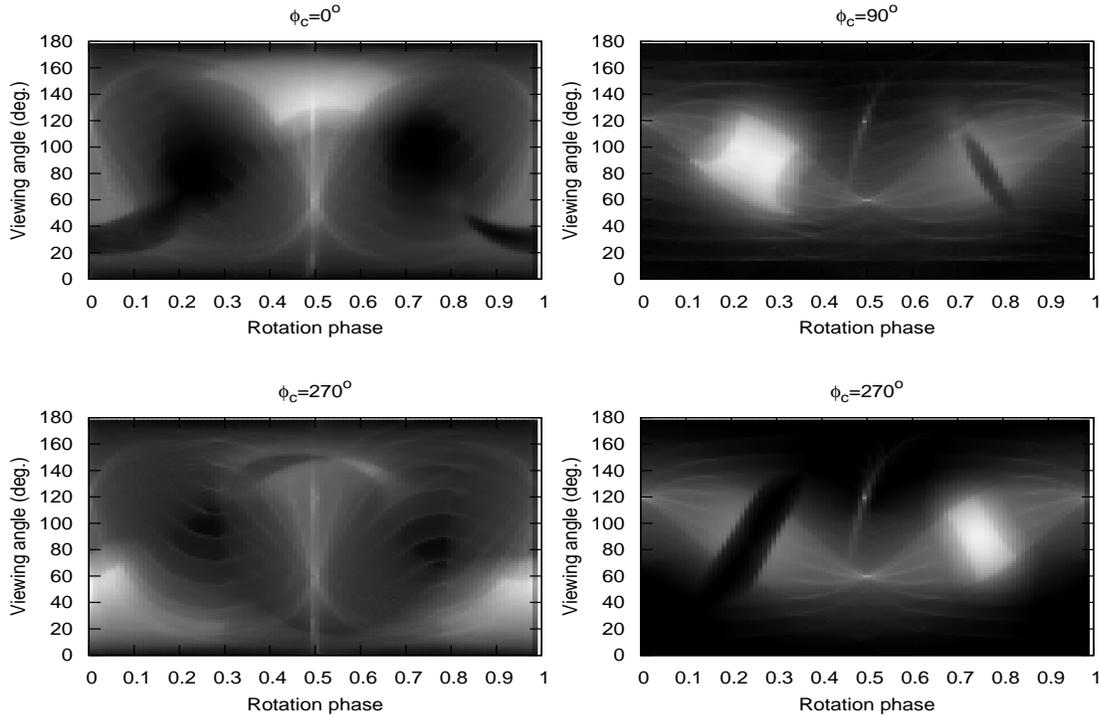}
\caption{Sky-maps of GeV $\gamma$-ray emissions
 for the inclination angle of $\alpha=60^{\circ}$ and $a=16$. 
The centre of the emitting magnetic flux tube  is $\phi_c=0^{\circ}$ (upper left), $90^{\circ}$ (upper right), 
$180^{\circ}$ (lower lift) and $270^{\circ}$ (lower right), respectively. The whiteness refers the emissivity of the radiation.}
\label{map}
\end{figure}

\begin{figure}
\includegraphics[height=12cm, width=12cm]{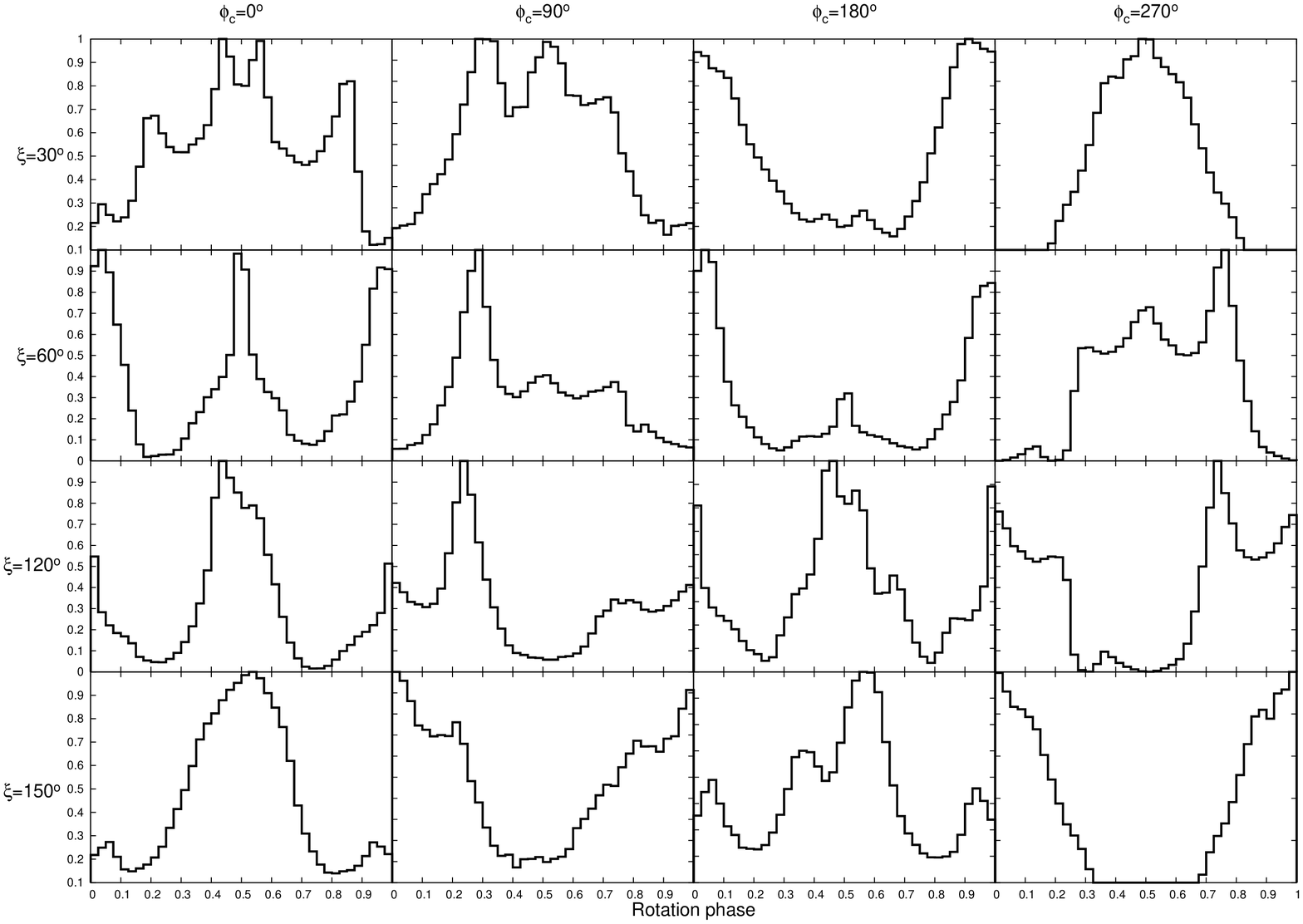}
\caption{The calculated light curves for  the inclination angle of $\alpha=60^{\circ}$ and $a=16$. From the top to bottom, 
the calculated light curves are for 
$\xi=30^{\circ}, 60^{\circ},120^{\circ}$ and $150^{\circ}$, respectively.  }
\label{light}
\end{figure}

The predicted light curves for the inclination angle $\alpha=60^{\circ}$ 
are summarized in Figures~\ref{map} and \ref{light}, which are the 
sky-photon-mapping and the calculated light curves, respectively. 
We assume that 
the azimuthal angle of  the centre of the emitting magnetic flux tube, which 
has a width $\delta\phi\sim1$~radian, 
 is $\phi_c=0^{\circ},90^{\circ}, 180^{\circ}$ and $270^{\circ}$. 
In Figure~\ref{map}, the whiteness refers the emissivity of the radiation
 on ($\xi$, $\Phi$)-plane, where $\xi$ and $\Phi$ are 
the Earth viewing angle and 
the rotation phase, respectively. The brighter region  (e.g. $\Phi\sim 0.2$ 
and $\xi\sim 100^{\circ}$ in the panel for $\phi_c=90^{\circ}$) corresponds 
to the rotation phase at which  both radiations produced by 
the particles migrating from north to south poles and from south
 to north poles contribute to the observed emissions, 
 while the dark region (e.g. 
$\Phi\sim 0.75$ and $\xi\sim 80^{\circ}$ in the panel for  $\phi_c=90^{\circ}$),
 both emissions passes through the region $r\le 5\times 10^{7}$~cm and do not 
contribute to the observed emissions. 

In Figure~\ref{light}, from top to bottom, the calculated light curves are for 
$\xi=30^{\circ}, 60^{\circ},120^{\circ}$ and $150^{\circ}$, respectively. 
 We can see in Figure~\ref{light} that  the calculated light curves 
have in general broad peak in one-rotation period with sometimes sharp   
peaks (e.g. $\phi_c=90^{\circ}$ and $\xi=120^{\circ}$).  
 For the emission from the closed field lines, 
the special relativistic effects can be ignored,
 while it becomes important to explain the sharp and narrow pulse profiles of 
 canonical $\gamma$-ray pulsars (Romani \&  Yadigaroglu 1995), in which 
the observed emissions are produced on the open field line region. 
In the present scenario, on 
the other hand, the sharp peak appears if the radiations from both  kinds of
 particles migrating toward south pole and toward north pole are observed simultaneously.

\subsection{Spectrum}
\label{spectrum}
To calculate the typical spectrum, we assume that the acceleration region 
 extends beyond  $r\ge 10^{7}$~cm and all of GeV $\gamma$-rays emitted 
above $r_c=5\times 10^7$~cm  can escape from the pair-creation process. 
Between $r_{c1}\sim 10^7$cm and  $r_c$,  furthermore, all synchrotron photons
 emitted by the outwardly migrating pairs, 
which were produced by outwardly propagating  $\gamma$-rays, 
 can escape from the pair-creation processes, 
as discussed in section~\ref{xemi}. The spectrum of the curvature radiation 
at each calculation grid  is expressed as
\begin{equation}
P_c(E_{\gamma})=\frac{\sqrt{3}e\Gamma_p}{hcR_{c}}F(\chi)j_a A,
\end{equation}
where $\chi=E_{\gamma}/E_c$ with $E_c=3hc\Gamma^3/4\pi R_c$ and 
\[
F(\chi)=\chi\int_{\chi}^{\infty}K_{5/3}(\xi)d\xi,
\]
where $K_{5/3}$ is the modified Bessel function of  order 5/3. In addition, 
$j_a$  is the current density and $A$ is the cross section of the grid perpendicular to the dipole field. In this section, we assume the 
Goldreich-Julian value, $j_a(r)=\Omega B_d(r)/2\pi$, for the current density. 
We assume that the Lorentz factor of the accelerated particles is given by 
equation~(\ref{gmax}).

The synchrotron radiation of the outwardly migrating 
 pairs produced between $r_{c1}$ and $r_c$ (c.f. section~\ref{xemi}) may be 
described by 
\begin{equation}
P_{s}(E_s)=\frac{\sqrt{3}e^3B\sin\theta_p}{m_ec^2h}
\int\frac{{\rm d}N_e}{{\rm d}E_e}F(\zeta){\rm d}E_e, 
\label{synchro}
\end{equation}
where $\zeta=E_s/E_{syn}$ with 
$E_{syn}=3\Gamma_{e\pm}^2ehB_d(r)\sin\theta_p/4\pi m_ec$ and $\theta_p$ the 
pitch angle. The distribution of the pairs under the steady condition 
may be described as 
\begin{equation}
\frac{dN_{e}}{dE_{e}}
\sim \frac{P_c(E_{\gamma})}{\dot{E}_{syn}},~~\Gamma_{e\pm,min}\le \Gamma_{e\pm}
\le E_{\gamma}/2m_ec^2, 
\end{equation}
where $\dot{E}_{syn}=2e^2B^2\sin^2\theta_p\Gamma_{e\pm}^2/3m^2c^5$ 
is the energy loss 
rate of the synchrotron radiation.  The minimum Lorentz factor 
 of the pairs 
is $\Gamma_{e\pm,min}\sim 1/\sin\theta_p\sim m_ec^2B_c/(\chi E_{\gamma}B_d(r))$.

\begin{figure}
\includegraphics[height=10cm, width=15cm]{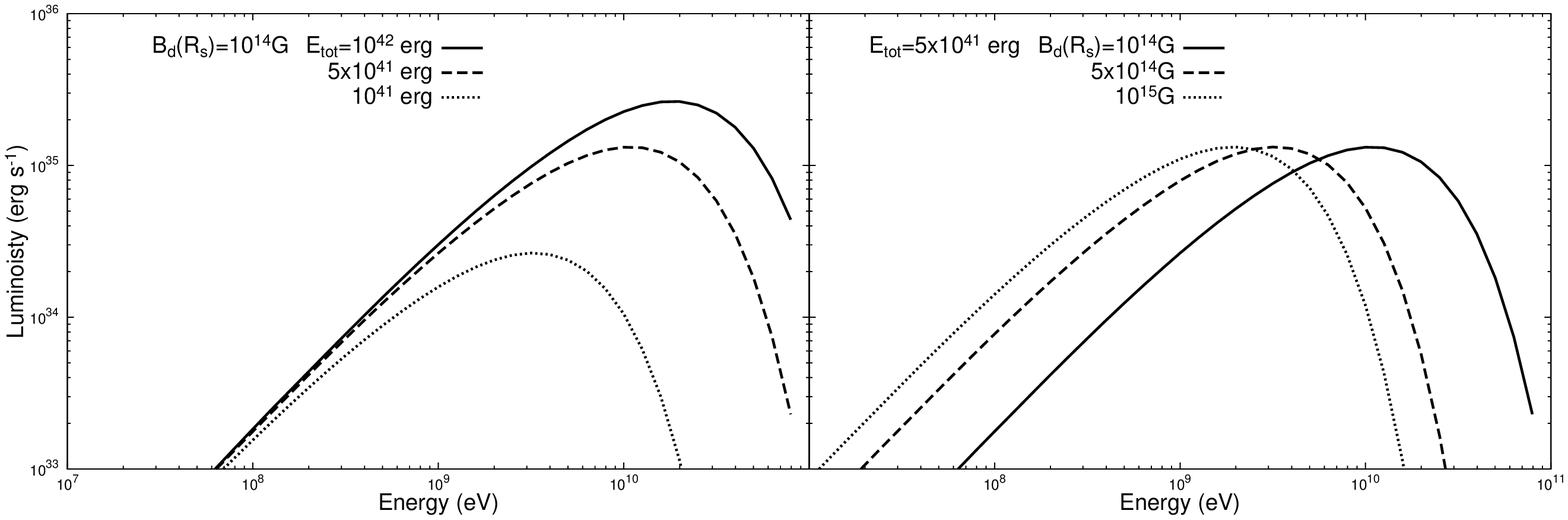}
\caption{Spectrum of the curvature radiation from the oscillating 
magnetic flux tube. The results are for $\alpha=60^{\circ}$,
 $\Omega=1{\rm s^{-1}}$, $\phi_c=0^{\circ}$ and $a=16$. The acceleration 
region extends beyond $r\ge 10^7$~cm and the all emissions above (below) 
$r=5\times 10^7$cm  can (cannot) escape from the pair-creation processes. 
Left: $B_d(R_s)=10^{14}$G. The solid, dashed and dotted 
lines are result for $E_{tot}=10^{42}{\rm erg}$, $5\times 10^{41}{\rm erg}$ 
and $10^{41}{\rm erg}$, respectively. Right: $E_{tot}=5\times 10^{41}$G. 
The solid, dashed and dotted 
lines are result for $B_d(R_s)=10^{14}{\rm G}$, $5\times 10^{14}
{\rm G}$  and $10^{15}{\rm G}$, respectively.}
\label{spec}
\end{figure}

Figure~\ref{spec}  summarizes typical spectra predicted by the
 present scenario.  In Figure~\ref{spec}, we have used 
  $\alpha=60^{\circ}$ for the inclination angle, 
 $a=16$ for the magnetic surface of the typical emission region,  and $\phi_c=0^{\circ}$ for 
the azimuthal angle of the centre of oscillating magnetic flux tube. 
 In Figure~\ref{spec}, the right and left panels show the dependency
 of the spectra of the curvature radiation  on the released 
energy ($E_{tot}$) and on the global dipole field ($B_d(R_s)$), respectively. 
The results are for $\Omega=1~{\rm s^{-1}}$ and $\ell=10^{5}$cm. 
We find in Figure~\ref{spec} that the luminosity is 
proportional to the released total energy because $L_{\rm \gamma}\propto 
\delta B^2(R_s)\propto E_{\rm tot}$, as indicated by equation~(\ref{power}).
 As we can see in Figure~\ref{spec},  the typical energy of the curvature 
spectrum increases with the released energy,  
that is, $E_c\propto \delta B^{3/2}(R_s)\propto  E_{tot}^{3/4}$ 
(c.f. equation~(\ref{cene})).  In the right panel of Figure~\ref{spec}, 
we see that the typical energy of the curvature radiation  
decreases with increase of the magnetic field. This is because 
the displacement of the footprints of the oscillating magnetic field lines
 becomes smaller for stronger background magnetic field,
 as equation~(\ref{dell}) indicates. The 
decrease of the displacement implies  the decrease  
of  the potential drop (equation~(\ref{poten})) and hence the decrease of 
 the typical energy of 
the curvature radiation, $E_c\propto B_d^{-3/4}(R_s)$ (equation~\ref{cene}). 

\begin{figure}
\includegraphics[height=10cm, width=10cm]{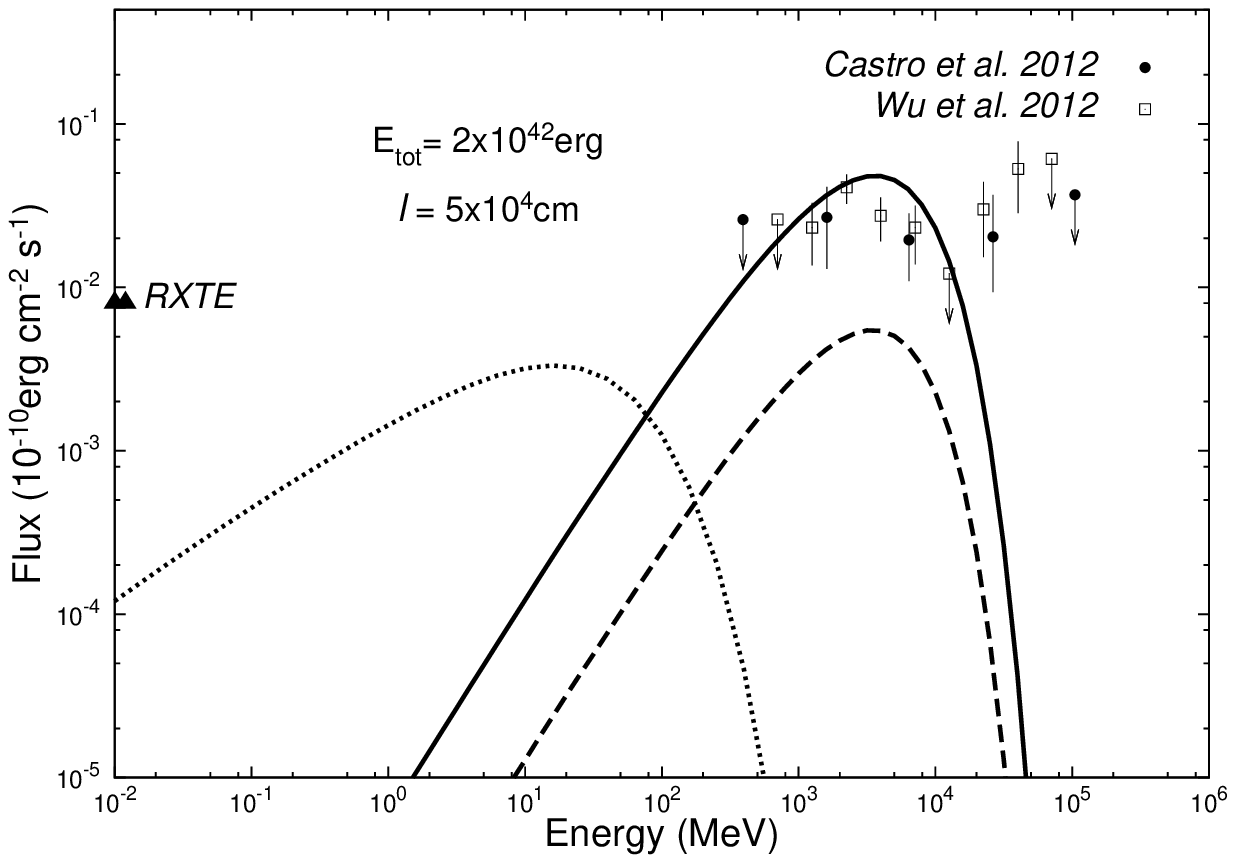}
\caption{The spectrum of  AXP 1E~2259+586. The solid and dashed lines 
are curvature radiation above and below  $r=5\times 10^7$~cm, respectively.
 The results are for $E_{tot}=2\times 10^{42}$erg and $\ell=5\times 10^4$cm. 
The dotted line is spectrum of the synchrotron radiation from the first 
generation of the pairs produced  $10^7{\rm cm}\le r\le 5\times 10^7{\rm cm}$, 
if all curvature photons above 1GeV  (dashed line)  are 
absorbed by magnetic field. The data are taken from Castro et al. (2012) and 
from Wu et al. (2012)  for the $Fermi$ and from Kuiper et al. (2006) 
for the $RXTE$, respectively.}
\label{spec3}
\end{figure}

In Figure~\ref{spec3}, we present the predicted spectrum in wide energy
 band using the parameters of AXP 1E~2259+586 ($\Omega\sim 0.9 {\rm s^{-1}}$ 
and $B_d(R_s)\sim 10^{14}$G). AXP 1E~22459+586 has shown frequent glitches, 
which sometimes accompany the X-ray outbursts. Furthermore, 
the GeV $\gamma$-ray 
emissions in the direction of CTB~109, which will be associated with AXP 
1E~22459+586, were founded in the $Fermi$ data (Castro et al. 2012). Although 
the origin from SNR has been suggested, the possibility of the 
emissions from AXP 1E~22459+586 has not been ruled out yet. 

The model spectrum  in Figure~\ref{spec3} is result for  
 $E_{tot}=2\times 10^{42}{\rm erg}$, $\ell=5\times 10^{4}$~cm, 
$\phi_c=0^{\circ}$ and  $a=13$. The solid line and  dashed line 
are spectra of the curvature radiation above $r\ge 5\times 10^{7}$cm and 
between $10^7{\rm cm}\le r\le 5\times 10^{7}$cm, respectively. The dotted line 
represents the synchrotron spectrum from the first generation of pairs, 
 if  all outgoing photons above 1GeV 
emitted between  $10^7{\rm cm}\le r\le 5\times 10^{7}$cm (dashed line) are 
absorbed by the magnetic field (c.f. section~\ref{xemi}).  
The measured spectrum 
is represented by the filled circles of Figure~\ref{spec3}. We can 
seen in Figure~\ref{spec3} that  if the observed 
GeV emissions would be originated from the magnetosphere, 
the observed flux level could be explained by the present scenario with  
the total released energy  of  $E_{tot}=2\times 10^{42}$erg 
 and the typical size of the cracking
 platelet of $\ell \sim 5\times 10^{4}$cm, respectively. 
 However, the present model can reproduce only  emissions 
below  $\sim$10~GeV, as Figure~\ref{spec3} shows.  Above 10~GeV, 
the present model expects that the emissions from SNR dominate 
the magnetospheric emissions.
 
The synchrotron radiation  
near $r_c\sim 5\times 10^7$~cm (dotted line) can extend down to hard X-ray 
bands, but its flux level is well  below  
the persistent X-ray emissions measured 
 by RXTE (filled triangles, Kuiper et al. 2006). Hence, it will be 
difficult to observe the synchrotron radiation in the outer magnetosphere.

\section{Discussion}
\label{discuss}
We have discussed possible scenario for the pulsed GeV $\gamma$-ray radiation 
in outer magnetosphere of the magnetar. The GeV $\gamma$-rays will escape from
 the pair-creation process if the emission process is occurred beyond $r_c\sim 
5\times 10^{5}$~cm. In the present scenario, the  Alfv$\acute{\rm e}$n wave  
carries the magnetic energy released by the crust cracking of the 
magnetic field into outer 
magnetosphere $r\ge r_c$ along the background magnetic field. 
The oscillation of the magnetic field induces the 
available potential drop $\delta\Phi_p\sim 10^{15}$~Volts, 
which can accelerate the electrons and/or positrons to the Lorentz
 factor $\Gamma\sim 10^{7}$. The curvature radiation at the outer
 magnetosphere can produce GeV $\gamma$-rays.

The pulsed GeV $\gamma$-ray radiation from the magnetars have not been reported
 yet, although the predicted luminosity 
$L_{\gamma}\sim 10^{35}~{\rm erg~s^{-1}}$ may be 
large enough to detect pulsed GeV $\gamma$-rays by the $Fermi$ telescope. 
 However, several reasons can be raised to explain the non-detection of 
the pulsed GeV emissions from the magnetars.  First, the magnetars are 
in general located at the Galactic plane, the background radiation may 
prevent the detection of the pulsed radiation. Second, the magnetars  have 
 shown  frequent glitches that are sudden changes in frequency 
and/or frequency derivative (\.{I}\c{c}dem et al. (2012)). 
Hence the timing parameters  of the magnetars are very unstable, 
which makes 
even harder to detect the pulsed period in the $Fermi$ data.  
Third, it has become clear that the magnetars are 
associated with  the SNRs (Allen \& Horvath, 2004; 
Gaensler et al. 2005; Halpern \& Gotthelf 2010). It is possible that 
 $\gamma$-ray emissions from SNRs dominate the pulsed emissions in the data. 
 Recent $Fermi$ observations have revealed  properties of 
the GeV radiations from  SNRs (Abdo et al. 2009, 2010, 2011). 
 For the young SNRs, the accelerated electrons at the forward shock 
may produce the GeV radiations through the 
inverse-Compton process, while for middle age SNRs, the interaction between the 
accelerated protons and ambient molecular clouds produces the bright 
GeV radiations via the $\pi^0$ decay process. 

We note that the typical luminosity of  GeV radiation of SNRs is order 
of $10^{34-35}~\mathrm{erg~s^{-1}}$, which is same order of magnitude predicted 
by the present magnetosphereic emission model.  Hence, it would be possible
  that the GeV emissions in the direction of magnetars 
are composed of the emissions from SNRs and magnetospheres.  
The pulsed radiation predicted by the present scenario 
 will change its luminosity level $L_{\gamma}\sim 10^{34-35}\mathrm{erg~s^{-1}}$
 at a temporal  scale of  years after the energy injection into 
the magnetosphere, while 
the SNR's emission will be stable. Hence,  a temporal behavior 
of the observed GeV emissions will discriminate between the two components.

Finally we note that it has been proposed that the glitch of normal
 pulsars is caused by unpinning of the super-fluid vortices from the lattice 
(Alpar 2001). The energy released by the glitch is deposited 
at inner crust, and most of energy is dissipated to heat up the entire star
 (Tang \& Cheng 2001). The increases of the surface X-ray emission may affect 
the gamma-ray emission because of the photon-photon pair-creation process.  
With a typical released energy $\sim10^{42}$erg, the glitch can 
increase surface X-ray emissions by a factor of 2-3 if the core temperature 
is $T_c\sim 10^7$K for old pulsars or only a few percent 
if $T_c\sim10^8$K for young pulsars (Tang \& Cheng 2001). Hence, 
it is expected such small change of the surface emission does
 not change much the $\gamma$-ray fluxes for young pulsars,
 e.g. Crab and Vela. For the magnetar case, on the other hand, 
the cracking of the outer crust could trigger the glitches that accompany 
the radiative outburst.  A part  of the released 
energy is carried into the magnetosphere via  Alfv$\acute{\rm e}$n waves, and  
 may be used  to accelerate the particles, which results in the GeV 
$\gamma$-ray emissions.


We express our appreciation to an anonymous referee for  useful  comments. 
We thank  M.Ruderman, S.Shibata, A.H. Kong, D.Hui and J.H.K. Wu, 
for the useful discussions.   This work was supported by a GRF grant of 
Hong Kong Government under HKU700911P.

\appendix
\begin{figure}
\includegraphics[height=7cm, width=10cm]{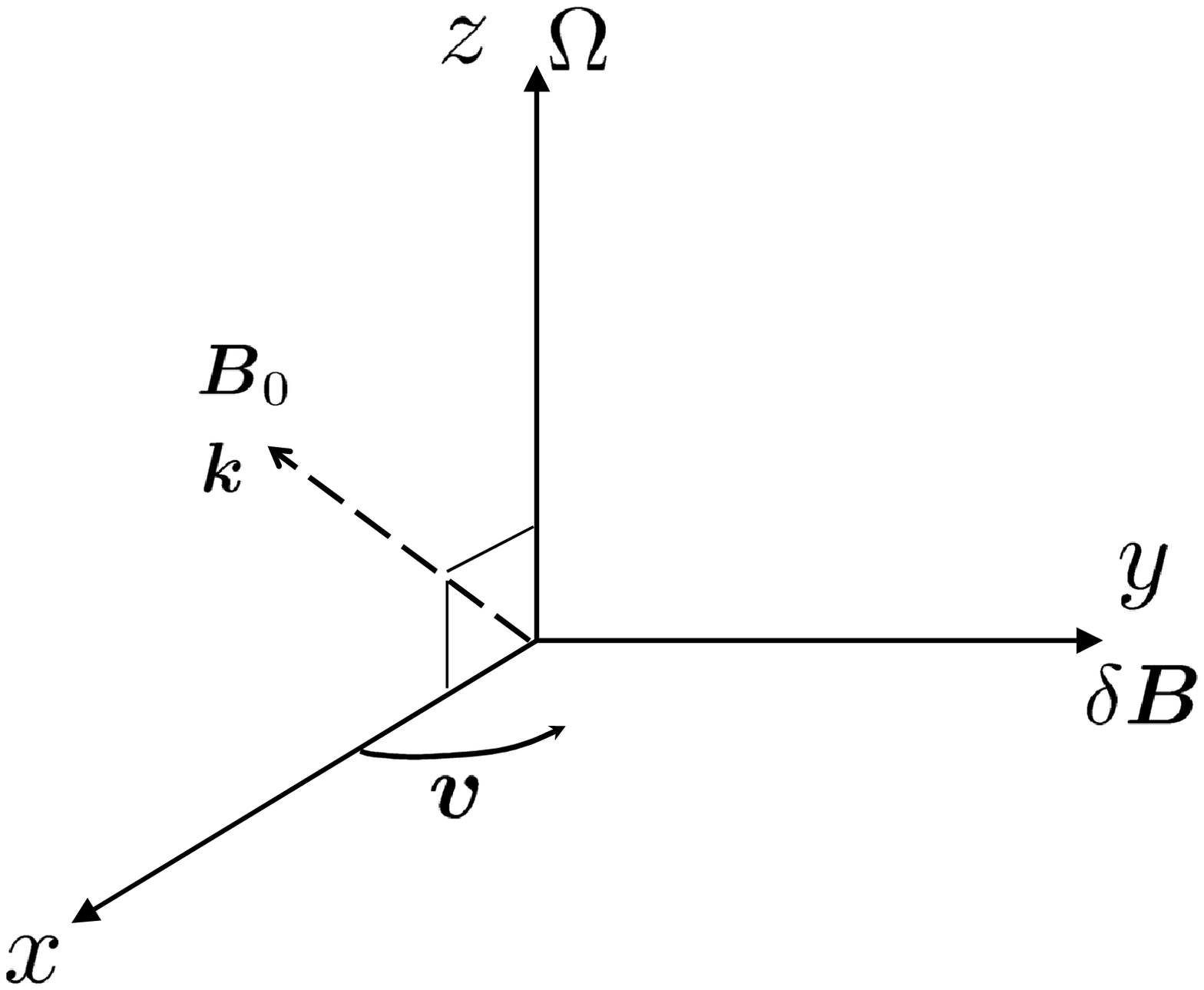}
\caption{Geometry of the Alfven wave propagating along the magnetic field line. The unperturbed fluid rotates around 
the spin axis, $\vec{v}=v_x\vec{e}_x+v_y\vec{e}_y$. }
\label{coor}
\end{figure}

\section{Propagation of  Alfv$\acute{\rm e}$n wave}
Within the frame work of a simple geometry of the co-rotating 
magnetosphere (c.f. Figure~\ref{coor}),
 we discuss the  Alfv$\acute{\rm e}$n wave 
propagating along the magnetic 
field line.  We define the z-axis as the spin axis, and assume an uniform 
magnetic field with  x and z components, 
\begin{equation}
\vec{B}_0=B_x\vec{e}_x+B_z\vec{e}_z.
\end{equation}
The co-rotation velocity is expressed as 
\begin{equation}
\vec{v}_0=v_x\vec{e}_x+v_y\vec{e}_y
\end{equation}
where $v_x=-\Omega y$ and $v_y=\Omega x$, respectively. 
We will assume  a uniform  perturbation of the magnetic field
\begin{equation}
 \delta\vec{B}=\delta B\vec{e}_y,
\end{equation}
where $\delta B$ is constant in both $\vec{x}$ and time $t$.
The linear differential equations for the equation of motion and the Maxwell equations are written as 
\begin{equation}
 \rho\left[\frac{\partial \delta \vec{v}}{\partial t}+(\vec{v}_0\cdot\nabla)\delta\vec{v}
+(\delta v\cdot\nabla)\vec{v}_0\right]=\delta \vec{F},
\label{eqm}
\end{equation}
\begin{equation}
 \nabla\cdot\delta\vec{E}=4\pi \delta\rho_e,
\end{equation}
\begin{equation}
 \nabla\cdot \delta \vec{B}=0,
\label{nabb}
\end{equation} 
\begin{equation}
\nabla\times \delta\vec {E}=-\frac{1}{c}\frac{\partial \delta \vec{B}}{\partial t},
\end{equation}
and 
\begin{equation}
\nabla\times\delta\vec{B}=\frac{4\pi}{c}\delta\vec{j}+\frac{1}{c}\frac{\partial \delta\vec{E}}{\partial t},
\end{equation}
where the perturbation of the Lorentz forces can be written down as
\begin{equation}
 \delta \vec{F}=\delta\rho_e\vec{E}+\rho_e\delta\vec{E}+\frac{1}{c}(\delta \vec{j}\times \vec{B}_0
+\vec{j}\times \delta\vec{B}),
\label{fro}
\end{equation}
where $\rho_e$ and $\vec{j}=\rho_e\vec{v}_0$ are background charge 
density and  current, respectively. 
The background charge density, $\rho_e$, is described  by the Goldreich-Julian 
charge density, that is, $\rho_e=\nabla\cdot\vec{E}/4\pi=-\nabla\cdot(\vec{v}_0\times\vec{B}_0)/4\pi\sim -\Omega B_z/2\pi c$.  The remaining equation is 
given by  the frozen in condition 
\begin{equation}
 \delta\vec{E}=-\frac{1}{c}(\delta\vec{v}\times \vec{B}_0+\vec{v}_0\times \delta\vec{B}).
\end{equation}

We may seek a solution in the form of the plane wave propagating along the background magnetic field;
\begin{equation}
 \vec{f}\propto {\rm exp}\{i[k(b_xx+b_zz)-\omega t]\},
\end{equation}
where  $b_x=B_x/B$, $b_z=B_z/B$, $\omega$ is the wave frequency and $k$ is 
 the magnitude of the wave number.  As this take place in 
the set of the linear equations (\ref{eqm})-(\ref{fro}), we 
can obtain the electric field as 
\begin{equation}
\delta E_x=-b_z(-\beta_{\phi}+b_x\beta_x)\delta B,~
\delta E_y=0,~
\delta E_z=-(b_x\beta_{\phi}+\beta_xb_z^2)\delta B,
\end{equation}
and the current as
\begin{equation}
\delta j_x=-i\frac{kc}{4\pi}(b_z\delta B-\beta_{\phi}\delta E_x),~
\delta j_y=0,~
\delta j_z=i\frac{kc}{4\pi}(b_x\delta B+\beta_{\phi}\delta E_z),
\end{equation}
where $\beta_{\phi}=\omega/kc$ is the phase velocity in units of the speed of light and $\beta_x=v_x/c$. 

The dispersion relation is given by 
\begin{equation}
 (b_x\beta_x-\beta_{\phi})^2-\frac{b_x^2\Omega^2}{k^2c^2}
+b_z^2\beta_x^2\beta_A^2-\beta_A^2(1-\beta_{\phi}^2)=0,
\end{equation}
where $\beta_A=v_A/c$ with $v_A=B_0/\sqrt{4\pi \rho}$ being the  Alfv$\acute{\rm e}$n velocity. Usually, the second term in the left hand side can be negligible because $\Omega/kc\sim \Omega /\omega \ll 1$. In the limit 
of $\beta_x\ll \beta_A$ and $\beta_y=v_y/c\ll \beta_A$,  the phase velocity becomes 
\begin{equation}
 \beta_{\phi}^2=\frac{\beta_A^2}{1+\beta_A^2}
\end{equation}
which is the traditional dispersion relation for the  Alfv$\acute{\rm e}$n wave, and we obtain  $\beta_{\phi}=1$ for $\beta_A>>1$ and 
$\beta_{\phi}=\beta_A$ for $\beta_A\ll 1$. 

The electric field parallel ($\delta E_{||}$) and perpendicular ($\delta E_{\perp}$)
 to  the background magnetic field can be described by
\begin{equation}
\delta E_{||}\equiv |\delta \vec{E}\cdot \vec{B}_0|=|b_z\beta_x\delta B|
\end{equation}
and 
\begin{equation}
\delta E_{\perp}\equiv |\delta \vec{E}\times \vec{B}_0|=|\beta_{\phi}\delta B|,
\end{equation}
respectively. For the current, we obtain
\begin{equation}
\delta j_{||}=-i\frac{\omega}{4\pi}b_z\beta_x\delta B,
\end{equation}
and
\begin{equation}
\delta j_{\perp}=i\frac{kc}{4\pi}(1-\beta_{\phi}^2)\delta B,
\end{equation}
respectively. The important result is that the electric field  and current along the magnetic field are 
 exited due to the rotation of the unperturbed matter, 
and hence the electrons and/or positrons can be accelerated 
along the magnetic field line. 

In the limit of $\beta_A\gg 1$ ($\beta_{\phi}=1$), we find that  
\begin{equation}
 (\delta E_{||},~\delta E_{\perp})
\sim -\delta B(b_z\beta_x,~1).  
\end{equation}
Hence the magnitude of the  induced electric field is in order of 
magnitude of  the perturbed magnetic field. The current becomes 
\begin{equation}
(\delta j_{||},~\delta j_{\perp})\sim i\frac{\omega}{4\pi}\delta B
(-b_z\beta_x,~\beta_A^{-2}), 
\end{equation}
implying the parallel current   dominates the perpendicular current. 

In the limit of $\beta_{x,y}\ll\beta_A\ll 1$ ($\beta_{\phi}=\beta_A$), the electric field and the current become
\begin{equation}
 (\delta E_{||},~\delta E_{\perp})\sim -\delta B (b_z\beta_x, \beta_{\phi})
\end{equation}
and 
\begin{equation}
  (\delta j_{||},~\delta j_{\perp})\sim  i\frac{\omega}{4\pi}\delta B
(-b_z\beta_x,~\beta_A^{-1}),
\end{equation}
respectively. The magnitude of induced electric field is much smaller than that of the perturbed magnetic field.  

Finally, if $\beta_A\sim \beta_{x,y}\ll 1$, we will see the phase velocity 
\begin{equation}
\beta_{\phi}\sim b_x\beta_x\pm \beta_{A}. 
\end{equation}


\begin{thebibliography}{}
\bibitem[\protect\citeauthoryear{Abdo}{2011}]{ab11}
Abdo et al. 2011, ApJ, 734, 28
\bibitem[\protect\citeauthoryear{Abdo}{2010}]{ab10}
Abdo et al. 2010, Sci, 327, 1103
\bibitem[\protect\citeauthoryear{Abdo}{2009}]{ab09}
Abdo et al. 2009, ApJL, 706, 1
\bibitem[\protect\citeauthoryear{Allen}{2004}]{al04}
Allen, M.P. \&  Horvath, J.E., 2004, ApJ, 616, 346
\bibitem[\protect\citeauthoryear{Alpar}{2001}]{al01}
Alpar, M.A., 2001, preprint (astro-ph/0112306)
\bibitem[\protect\citeauthoryear{Alpar}{1984}]{al84}
Alpar, M.A.,  Langer, S.A. \&  Sauls, J. A., 1984, ApJ, 282, 533
\bibitem[\protect\citeauthoryear{Baring}{2007}]{Ba07}
Baring, M.G., \& Harding, A.K., 2007, Ap\&SS, 308, 109 
\bibitem[\protect\citeauthoryear{Beloborodov}{2012}]{Be12}
Beloborodov, A.M., 2012, ApJ, submitted (arXiv1201.0664B)
\bibitem[\protect\citeauthoryear{Beloborodov}{2007}]{Be07}
Beloborodov, A.M. \& Thompson,C., 2007, ApJ, 657, 967
\bibitem[\protect\citeauthoryear{Beskin}{2000}]{Be00}
Beskin, V.S. \& Rafikov, R.R., 2000, MNRAS, 313, 445
\bibitem[\protect\citeauthoryear{Blaes}{1989}]{B89}
Blaes, O., Blandford, R., Goldreich, P. \&  Madau, P., 1989, ApJ, 343, 839B
\bibitem[\protect\citeauthoryear{Castro}{2012}]{ca12}
Castro, D., Slane, P., Ellison, D.C., \& Patnaude, D.J., 2012, eprint 
arXiv:1207.1432
\bibitem[\protect\citeauthoryear{Cheng}{2001}]{ch01}
Cheng, K. S. \& Zhang, L., 2001, ApJ, 562, 918
\bibitem[\protect\citeauthoryear{Cheng}{1998}]{ch98}
Cheng, K. S., Gil, J. \& Zhang, L., 1998, ApJL, 493, 35
\bibitem[\protect\citeauthoryear{Dib}{2012}]{Di12}
Dib, R., Kaspi, V.M., Scholz, P., \&  Gavriil, F.P., 2012, ApJ, 748, 3
\bibitem[\protect\citeauthoryear{Enoto}{2010}]{en10a}
Enoto, T. et al., 2010, ApJ, 715, 665
\bibitem[\protect\citeauthoryear{Fatuzzo}{1993}]{fa93}
Fatuzzo, M. \& Melia, F. 1993, ApJ, 407, 680
\bibitem[\protect\citeauthoryear{Fernandez}{2011}]{fa11}
Fern\'{a}ndez, R. \& Thompson, C., 2007, ApJ, 660, 615
\bibitem[\protect\citeauthoryear{Gaensler}{2005}]{ga05}	
Gaensler, B.M., McClure-Griffiths, N.M.,  Oey, M.S.,  Haverkorn, M., 
 Dickey, J. M. \&  Green, A. J., 2005, ApJL, 620, 95
\bibitem[\protect\citeauthoryear{Hartog}{2008}]{ha08}
den Hartog, P.R., Kuiper, L., Hermsen,W., Kaspi, V.M., Dib,R., 
Kn$\mathrm{\ddot{o}}$dlseder, J., \&  Gavriil, F.P., 2008, A\&A, 489, 245
\bibitem[\protect\citeauthoryear{Halpern}{2010}]{ha10}
Halpern, J.P., \&  Gotthelf, E.V., 2010, ApJ, 725, 1384
\bibitem[\protect\citeauthoryear{Harding}{2006}]{ha06}
Harding, A. K., \& Lai, D. 2006, Rep. Prog. Phys., 69, 2631
\bibitem[\protect\citeauthoryear{Icem}{2012}]{ic06}
 \.{I}\c{c}dem, A.,  Baykal, A., \& .{I}nam, \c{C},S., 
2012, MNRAS, 419, 3109
\bibitem[\protect\citeauthoryear{Kaspi}{2007}]{ka07}
Kaspi V. M., 2007, Ap\&SS, 308, 1
\bibitem[\protect\citeauthoryear{Kojima}{2004}]{ko04}
Kojima, Y. \& Okita, T., 2004, ApJ, 614, 922
\bibitem[\protect\citeauthoryear{Kuiper}{2006}]{ku06}
Kuiper, L., Hermsen, W., den Hartog, P.R. \&  Collmar, W., 2006, ApJ, 645, 556
\bibitem[\protect\citeauthoryear{Mereghetii}{2008}]{me08}
Mereghetti, S. 2008, A\&AR, 15, 225
\bibitem[\protect\citeauthoryear{Pons}{2012}]{po12}
Pons, J.A., \& Rea., N., 2012, ApJL, 750, 6
\bibitem[\protect\citeauthoryear{Rea}{2008}]{re08}
Rea, N., Zane, S., Turolla, R.,  Lyutikov, M. \& G$\ddot{\rm o}$tz, D., 
 2008, 686, 1245
\bibitem[\protect\citeauthoryear{Romani}{1995}]{ro95}
Romani, R.W. \&  Yadigaroglu, I.-A., 1995, ApJ, 438, 314
\bibitem[\protect\citeauthoryear{Ruderman}{1975}]{ru75}
Ruderman, M., \&  Sutherland, P.G. 1975, ApJ, 196, 51
\bibitem[\protect\citeauthoryear{Ruderman}{1991}]{ru91}
Rumerman, M., 1991, ApJ, 382, 587
\bibitem[\protect\citeauthoryear{Scholz}{2011}]{sc11}
Scholz, P., \&  Kaspi, V.M., 2011,  ApJ, 739, 94
\bibitem[\protect\citeauthoryear{Takata}{2007}]{Ta07}
Takata, J., Chang, H.-K. \& Cheng, K.S., 2007, ApJ, 656, 1044
\bibitem[\protect\citeauthoryear{Tang}{2001}]{Ta01}
Tang, Anisia P.S. \& Cheng, K.S., 2001, ApJ, 549, 1039
\bibitem[\protect\citeauthoryear{Thompson}{1995}]{Th95}
Thompson, C., \& Duncan, R.C., 1995, MNRAS, 275, 255
\bibitem[\protect\citeauthoryear{Thompson}{2002}]{Th02}
Thompson, C., Lyutikov, M. \&  Kulkarni, S.R., 2002, 574, 332 
\bibitem[\protect\citeauthoryear{Thompson}{2006}]{Th06}
Thompson, C.,  2006, ApJ, 651, 333
\bibitem[\protect\citeauthoryear{Wang}{2010}]{wa10}
Wang, Y., Takata, J. \&  Cheng, K.S., 2010, ApJ, 720, 178
\bibitem[\protect\citeauthoryear{Woods}{2006}]{wo06}
Woods, P. M., \& Thompson, C. 2006, in Compact Stellar X-ray Sources , ed. W.
H. G. Lewin \& M. van der Klis (Cambridge: Cambridge Univ. Press), 547
\bibitem[\protect\citeauthoryear{Woods}{2005}]{wo05}
Woods, P.M., et al., 2005, ApJ, 629, 985
\bibitem[\protect\citeauthoryear{Woods}{2004}]{wo04}
Woods, P.M. et al., 2004, ApJ, 605, 378
\bibitem[\protect\citeauthoryear{Wu}{2012}]{wu12}
Wu, J.H.K, et al., 2012 submitted
\bibitem[\protect\citeauthoryear{zhu}{2008}]{zh08}
Zhu, W., Kaspi, V.M., Dib, R., Woods, P.M., Gavriil, F.P., \&  Archibald, A.M., 
2008, ApJ, 686, 520


\end{thebibliography}
\end{document}